\documentclass[journal]{IEEEtran}
\ifCLASSINFOpdf
\else
\fi
\usepackage{cite}
\usepackage{amsmath,amssymb,amsfonts}
\usepackage{algorithmic}
\usepackage{graphicx}
\usepackage{textcomp}
\usepackage{xcolor}
\usepackage{threeparttable}
\usepackage{multirow}
\usepackage{subfigure}
\usepackage{algorithm}
\usepackage{mathrsfs}

\usepackage{array}
\newcommand{\PreserveBackslash}[1]{\let\temp=\\#1\let\\=\temp}
\newcolumntype{C}[1]{>{\PreserveBackslash\centering}p{#1}}
\newcolumntype{R}[1]{>{\PreserveBackslash\raggedleft}p{#1}}
\newcolumntype{L}[1]{>{\PreserveBackslash\raggedright}p{#1}}
\begin{document}

%
\title{Boundary-aware Context Neural Network for Medical Image Segmentation}
%
%
%

\author{Ruxin~Wang, Shuyuan~Chen, Chaojie~Ji, Jianping~Fan
        and~Ye~Li,~\IEEEmembership{Senior Member,~IEEE}
\thanks{This work was funded by National Natural Science Foundation of China (U1913210, 61771465), Major Special Project of Guangdong Province (2017B030308007), Strategic Priority CAS Project XDB38000000, Shenzhen Basic Research Projects (JCYJ20180703145202065), Shenzhen Science and Technology Innovation Project (JSGG20170823144843046).}
\thanks{Ye Li and Jianping Fan are the \emph{Corresponding author}.}
\thanks{Ruxin Wang is with Joint Engineering Research Center for Health Big Data Intelligent Analysis Technology, Shenzhen Institutes of Advanced Technology, Chinese Academy of Sciences, Shenzhen, China. (e-mail: rx.wang@siat.ac.cn)}
\thanks{Shuyuan Chen is with the School of Software Engineering, University of Science and Technology of China, Hefei, and Shenzhen Institutes of Advanced Technology, Chinese Academy of Sciences, Shenzhen, China. (e-mail: sa517020@mail.ustc.edu.cn)}
\thanks{Chaojie Ji is with Shenzhen Institutes of Advanced Technology, Chinese Academy of Sciences, Shenzhen, China. (e-mail: cj.ji@siat.ac.cn)}
\thanks{Jianping Fan is with the Shenzhen Institutes of Advanced Technology, Chinese Academy of Sciences, Shenzhen, China. (e-mail: jp.fan@siat.ac.cn)}
\thanks{Ye Li is with Joint Engineering Research Center for Health Big Data Intelligent Analysis Technology, Shenzhen Institutes of Advanced Technology, Chinese Academy of Sciences, Shenzhen, China. (e-mail: ye.li@siat.ac.cn)}}

%
%

\markboth{Journal of \LaTeX\ Class Files,~Vol.~14, No.~8, August~2015}%
{Shell \MakeLowercase{\textit{et al.}}: Bare Demo of IEEEtran.cls for IEEE Journals}
%

\maketitle

\begin{abstract}
	
Medical image segmentation can provide a reliable basis for further clinical analysis and disease diagnosis. The performance of medical image segmentation has been significantly advanced with the convolutional neural networks (CNNs). However, most existing CNNs-based methods often produce unsatisfactory segmentation mask without accurate object boundaries. This is caused by the limited context information and inadequate discriminative feature maps after consecutive pooling and convolution operations. In that the medical image is characterized by the high intra-class variation, inter-class indistinction and noise, extracting powerful context and aggregating discriminative features for fine-grained segmentation are still challenging today. In this paper, we formulate a boundary-aware context neural network (BA-Net) for 2D medical image segmentation to capture richer context and preserve fine spatial information. BA-Net adopts encoder-decoder architecture. In each stage of encoder network, pyramid edge extraction module is proposed for obtaining edge information with multiple granularities firstly. Then we design a mini multi-task learning module for jointly learning to segment object masks and detect lesion boundaries. In particular, a new interactive attention is proposed to bridge two tasks for achieving information complementarity between different tasks, which effectively leverages the boundary information for offering a strong cue to better segmentation prediction. At last, a cross feature fusion module aims to selectively aggregate multi-level features from the whole encoder network. By cascaded three modules, richer context and fine-grain features of each stage are encoded. Extensive experiments on five datasets show that the proposed BA-Net outperforms state-of-the-art approaches. 

\end{abstract}


\IEEEpeerreviewmaketitle

\section{INTRODUCTION}

\IEEEPARstart{I}{mage} segmentation plays an important role in medical image analysis, which aims to address pixel-wise and fine-grained lesion recognition \cite{r1,r2}. With the development and popularization of medical imaging technology and equipment, ultrasound, magnetic resonance imaging (MRI), computed tomography (CT) and other imaging modalities provide an intuitive and effective way to diagnose and scan different kinds of diseases. These techniques have been widely used in daily clinical research and treatment planning. For different types of clinical applications, segmentation has been adopted as a key step of image analysis, such as lung segmentation in CT images \cite{r3}, skin lesion segmentation in dermoscopy images \cite{r4}, colorectal cancer segmentation in endoscopy images \cite{r5} and cell segmentation in microscopy images \cite{r6}. Accurate lesion detection is critical to provide a reliable basis for further clinical analysis \cite{r7}, disease diagnosis \cite{r8}, therapy planning \cite{r9} and prognosis evaluation \cite{r10}. High precision is typically required in lesion segmentation which need to segment the focused parts and extract relevant features accurately \cite{r11}.

\begin{figure}[!t]
	\centering
	\subfigure{\includegraphics[width=2.3cm,height=2.0cm]{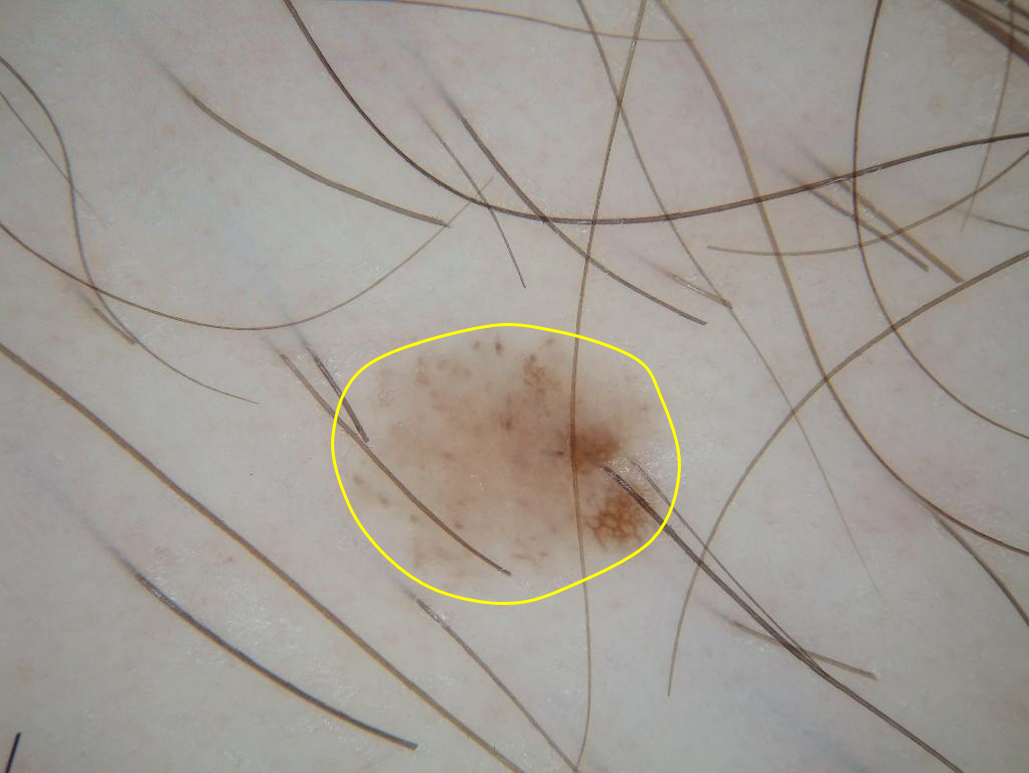}}
	\subfigure{\includegraphics[width=2.3cm,height=2.0cm]{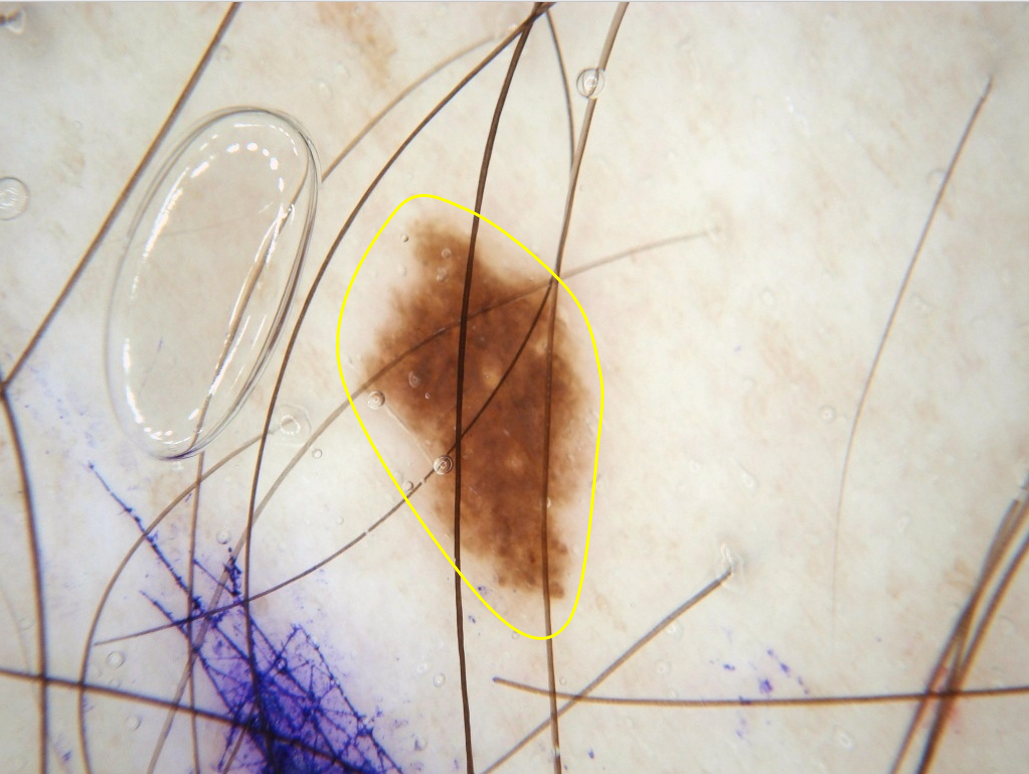}}
	\subfigure{\includegraphics[width=2.3cm,height=2.0cm]{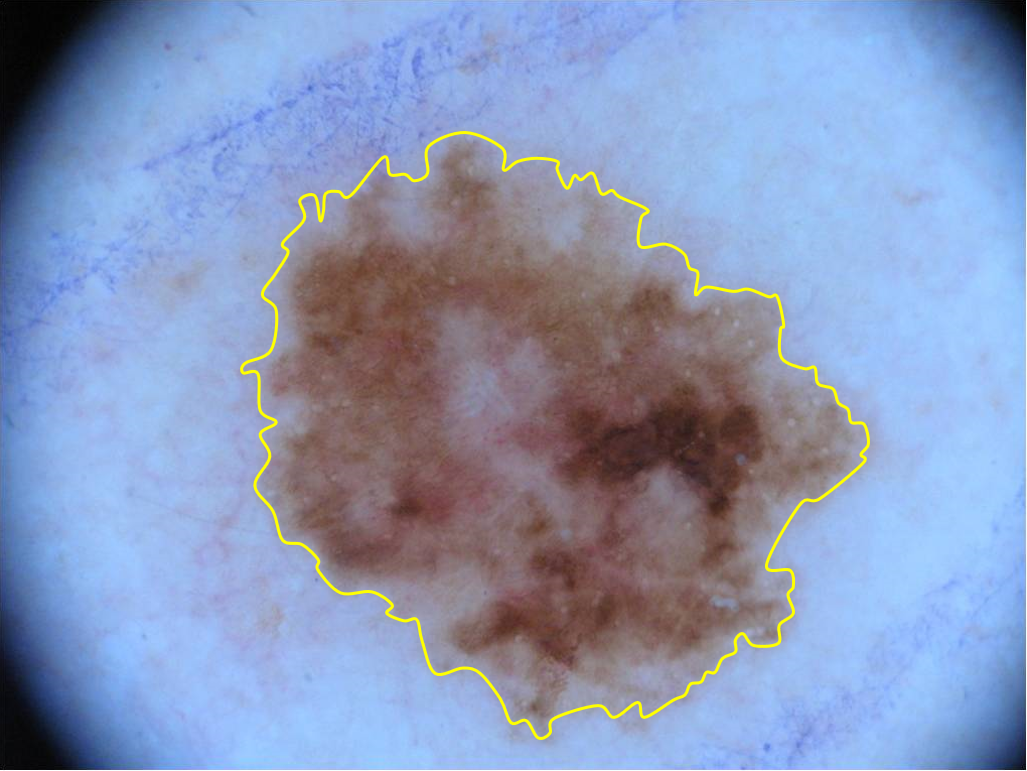}}
	\subfigure{\includegraphics[width=2.3cm,height=2.0cm]{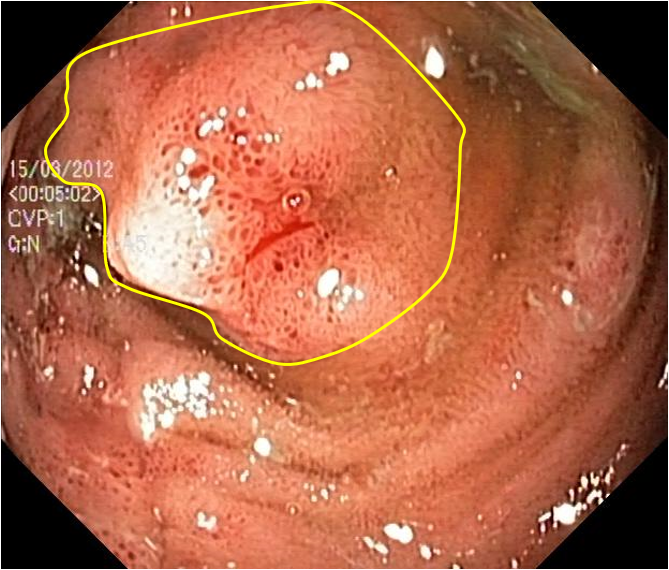}}
	\subfigure{\includegraphics[width=2.3cm,height=2.0cm]{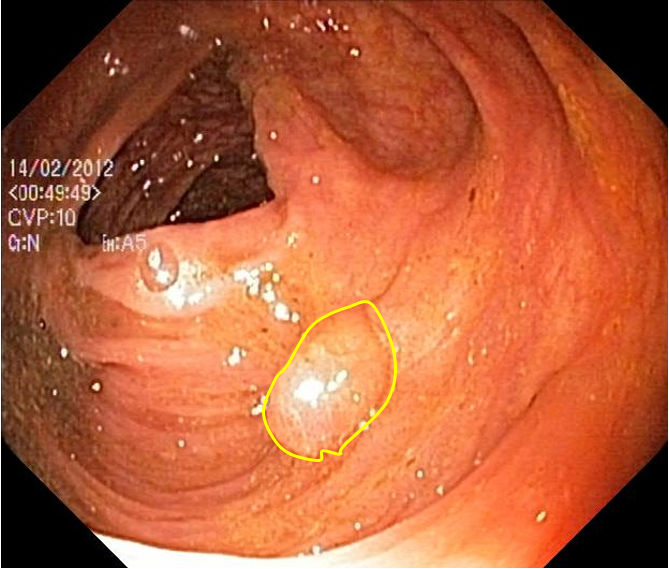}}
	\subfigure{\includegraphics[width=2.3cm,height=2.0cm]{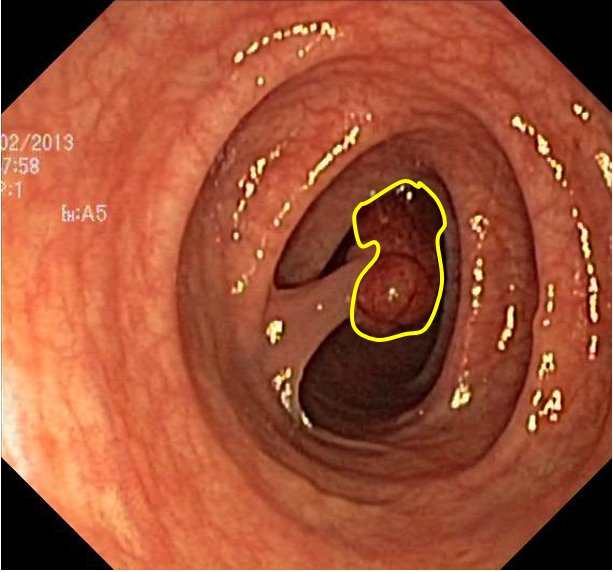}}
	\caption{Examples of two representative medical images. The first row shows the skin lesion in the dermoscopy image and the second row indicates the colorectal polyp in endoscopy images. In each image, yellow solid line refers to the target boundary.}
	\label{fig:1}
\end{figure}

With the increasing number of medical images and the development of Artificial Intelligence (AI), computer-assisted diagnosis technology can effectively assist professional clinicians to improve the accuracy and efficiency of analysis. However, automatic lesion (organ or tissue) recognition in medical image remains a complex and challenging task \cite{r12,r13}. At first, lesion regions have various sizes and shapes for different individuals. For some diseases, obvious individual differences increase the difficulty of recognition. Fig. \ref{fig:1} shows two examples for skin lesion and colorectal polyp. Secondly, The low contrast between the lesions and background also brings great challenge to segmentation. In particular, the focused area usually contains complex tissues and organs, which makes it very difficult to distinguish these confusing boundary pixels. In addition, some artifacts and imaging noise also impede the accuracy of segmentation. 

In the past few decades, a large group of automatic analysis algorithms for medical image segmentation have been proposed \cite{r14,r15,r16,r17}, which can be roughly classified into three categories: gray level-based \cite{r18}, texture-based \cite{r19} and atlas-based methods \cite{r20}. Although these approaches have enhanced the performance of automated segmentation through extracting different kinds of pixel and region features, they still have some common defects: 1) Traditional methods often design low-level hand-crafted features and make heuristic hypothesizes, which usually restricts the performance of prediction with complex scenarios. Moreover, abundant available information in the original image is neglected. 2) The robustness to artifacts, image quality and intensity inhomogeneity is low, which heavily depends on effective pre-processing. 

Recently, due to the remarkable successes of deep learning in computer vision, deep convolutional neural network (CNNs) has emerged as a promising alternative for medical image segmentation \cite{r21,r22,r23}, which successfully overcomes the limits of traditional hand-crafted features. Most state-of-the-art medical image segmentation approaches are based on encoder-decoder network architecture, among which the most representative methods are U-net \cite{r24} and fully convolutional network (FCN) \cite{r25}. The network framework is designed in an end-to-end way with pixel-wise supervision. In encoder stage, input image obtains high-level semantic feature representations through consecutive convolution operations. Then the top features of encoder network are employed to generate the predicted segmentation mask by progressive upsampling (uppooling or deconvolution) method in decoder network. Although convolutional neural networks have shown their advantages in the medical image segmentation task, most existing CNNs-based methods still suffer from inaccurate object boundaries and unsatisfactory segmentation results. This is caused by the limited context information and inadequate discriminative feature maps after consecutive pooling and convolution operations. In order to accurately recognize object, it is necessary to extract and aggregate high-level semantic features with low-level fine details simultaneously. Overall, how to learn richer context is still a challenge of segmentation algorithm for improving the recognition performance. 

Inspired by above analysis, we propose a novel convolutional network framework based on boundary-aware (BA-Net) for medical image segmentation, which follows the classical encoder-decoder structure. Specifically, in each stage of encoder network, pyramid edge feature extraction module (PEE) is proposed for obtaining edge information with multiple granularities firstly. Object boundaries define the shape of the object, and thus provide complementary cues for segmenting target objects. For getting richer knowledge about the sample, in each stage of encoder network, we design a mini multi-task learning module (mini-MTL) and jointly supervise segmentation and boundary map prediction during training. Furthermore, to take full advantage of features from different task, an interactive attention (IA) is proposed. IA makes use of the interactive information from different tasks to supervise the modeling of the target area recognition which is helpful to refine the segmentation performance. At last, a cross feature fusion module (CFF) is presented by selectively aggregate multi-level features from the whole encoder network, which further captures valuable context and preserves fine spatial information. By cascaded three modules, richer context and fine-grain features of each stage are encoded. In decoder network, we integrate these feature maps to get the final segmentation prediction sequentially. Finally, we evaluate our BA-Net on the multiple public medical image datasets and achieve consistent performance improvements on them. 

In summary, the contributions of this work are four-fold: 

1) We put forward a novel boundary-aware context neural network for 2D medical image segmentation, which employs PEE, mini-MTL and CFF modules to produce richer contextual information for guiding the decoder processing.

2) We design the PEE module and mini-MTL module with embedded interactive attention for fully mining the contextual features at the same level, which effectively leverage the boundary information for offering a strong cue to better segmentation prediction.

3) We build a CFF module to selectively incorporate cross-level features from other stage of the encoder network into current stage. In this way, information complementation among different levels of features is realized.

4) We conduct comprehensive experiments and achieve outstanding state-of-the-art segmentation performance for different tasks including skin lesion segmentation, colorectal polyp segmentation, lung segmentation and optic disc segmentation. The experimental results convince the efficiency of the proposed method.

The remainder of this paper is organized as follows: Section \uppercase\expandafter{\romannumeral2} reviews the recent development of medical image segmentation. Section \uppercase\expandafter{\romannumeral3} presents our proposed segmentation neural network in detail. In Section \uppercase\expandafter{\romannumeral4}, we evaluate the proposed model, and compare it with the state-of-the-art methods on the multiple public datasets. And related analysis of our method is discussed. Finally, Section \uppercase\expandafter{\romannumeral5} concludes this paper and prospects some future works.

\begin{figure}[!t]
	\centerline{\includegraphics[width=0.5\textwidth]{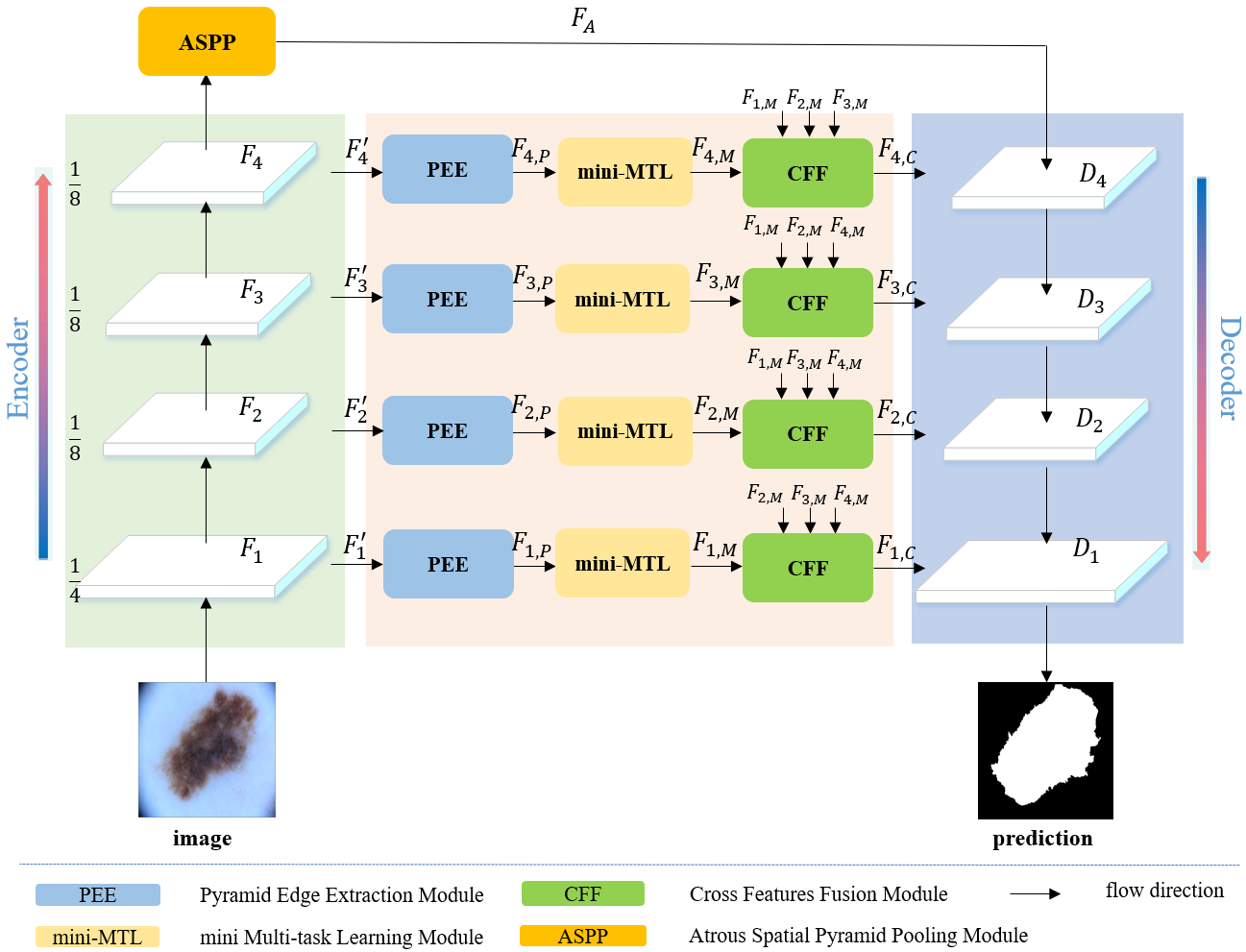}}
	\caption{The framework of the proposed model. In each stage of encoder network, we firstly obtain multiple granularities edge features by pyramid edge extraction module (PEE). Then the mini multi-task learning module (mini-MTL) jointly learns to segment object masks and detect lesion boundaries. At last, multi-level features from the whole encoder network are selectively aggregated by cross feature fusion module (CFF), which further captures valuable context and preserves fine spatial information.}
	\label{fig:2}
\end{figure} 

\section{RELATED WORK}

In this section, related conventional and CNN-based segmentation methods for medical image segmentation are briefly reviewed. 

\subsection{Conventional Segmentation Methods}

Over the past decades, various models have been introduced including gray level-based, texture-based and atlas-based models. Early methods based on gray level features mainly contain histogram statistic, edge detection and region growing strategies. For example, Carballido-Gamio et al. \cite{r26} applied the Normalized Cuts with local histograms of brightness to segment vertebral bodies from sagittal T1-weighted magnetic resonance images of the spine. Chung et al. \cite{r27} presented a partial-differential equations-based framework for detecting the boundary of skin lesions in dermoscopy images, which the object area is segmented either by the geodesic active contours or the geodesic edge tracing model. Nguyen et al. \cite{r28} proposed the watersnakes model for image segmentation by adding the contour length to the energy function. It integrates the strengths of watershed segmentation and energy based segmentation. Xie et al. \cite{r29} presented a novel texture and shape priors extracted by applying a bank of Gabor filters for kidney segmentation in ultrasound images. In Bazin's work \cite{r30}, they designed a segmentation framework based on both topological information and statistical atlases of brain anatomy which constrained topological equivalence between the prediction and the atlas. Although these models such as thresholding and region growing approaches are able to implement, but the performance is restricted in that the selection of threshold and region division criteria is greatly affected by the image intensity or texture information. Also hand-craft features for the segmentation are heavily relied on the experience of the researchers.

\subsection{CNN-based Segmentation Frameworks}

In recent years, deep convolutional neural networks have been successfully applied to a wide variety of problems in computer vision, which shows striking profit using encoder-decoder framework for image segmentation tasks \cite{r31,r32,r33}. In encoder network, image content is encoded by multiple convolutional layers from low-level to high-level. And in decoder part, the prediction mask is obtained by multiple upsampling (uppooling or deconvolutioanl) layers. In particular, image feature representation and context extraction are very crucial for segmentation task. For example, Chen et al. \cite{r34} proposed a segmentation framework named DeepLab which tailored an atrous spatial pyramid pooling module to encode the multi-scale contextual information by parallel dilated convolution. Zhao et al. \cite{r35} employed a pyramid pooling module with multiple pooling scales to capture multi-scale context in encoder network. Li et al. \cite{r36} presented a new dense deconvolutional network for skin lesion segmentation based on residual learning. It captures fine-grained multi-scale features of image for segmentation task by dense deconvolutional layers, chained residual pooling and auxiliary supervision. Xue et al. \cite{r37} proposed a end-to-end adversarial critic neural network with a multi-scale L1 loss function for medical image segmentation, which forces to learn both global and local features. Zhou et al. \cite{r38} tailored a collaborative network architecture to jointly improve the performance of disease classification and lesion segmentation by semi-supervised learning with attention mechanism. Chen et al. \cite{r39} proposed an unsupervised domain adaptation method to effectively tackle the problem of domain shift and achieved cross-modality image segmentation. Overall, effective extraction of image context is important for improving segmentation performance.

\section{METHODOLOGY}

In this section, we describe the construction of the proposed boundary-aware context neural network and the design methods of the three core modules (i.e. PEE, mini-MTL and CFF). Details of the proposed method are introduced as follows.

\subsection{Overview}

As shown in Fig. \ref{fig:2}, the proposed BA-Net has an encoder-decoder architecture design and starts with ResNet \cite{r40} as backbone (pre-trained on ImageNet \cite{r41}). In encoder network, the last global pooling and fully-connected layers of ResNet are removed firstly which only retains one convolution and four residual convolution blocks for primary features extraction. Without loss of generality, for an input image, 
we denote the output features of four residual blocks as $F_i, i\in\{1,2,3,4\}$. To further enlarge the receptive fields, the last two blocks in ResNet are modified using atrous convolution (atrous rate $= 2$) and maintain the same spatial resolution as the previous block by removing the pooling operation in our work. Thus, the output sizes of each block are $1/4$, $1/8$, $1/8$ and $1/8$ of the input image. In addition, an atrous spatial pyramid pooling (ASPP) module \cite{r34} is employed on the top feature maps of last residual block for capturing and encoding multi-scale features. The ASPP module comprises of four parallel atrous convolutions with different atrous rates and one global average pooling. The output features of ASPP are concatenated by upsampling and one $1\times 1$ convolution (with 256 channels), which further integrates and compresses the feature maps. In order to produce richer contextual information for guiding the decoder processing, we tailor three modules to fully mine the features in same level and aggregate other features from different levels in each stage of encoder network. Pyramid edge extraction module (PEE) is proposed for aggregating boundary information with multiple granularities of current level firstly. Then a mini multi-task learning module (mini-MTL) is adopted for getting richer knowledge leveraging potential correlations and complementary features among related boundary detection and segmentation tasks. At last, the learned feature maps of mini-MTL further module realize the complementarity among different levels and refine high-level features and low-level features through cross feature fusion module (CFF). Finally, in decoder part, we obtain the decoding features $D_i, i\in\{1,2,3,4\}$ by aggregating the output feature maps from the ASPP module and encoding features of each stage in turn for final segmentation prediction.
 
\subsection{Pyramid Edge Extraction Module}

\begin{figure}[htbp]
	\centerline{\includegraphics[width=0.45\textwidth]{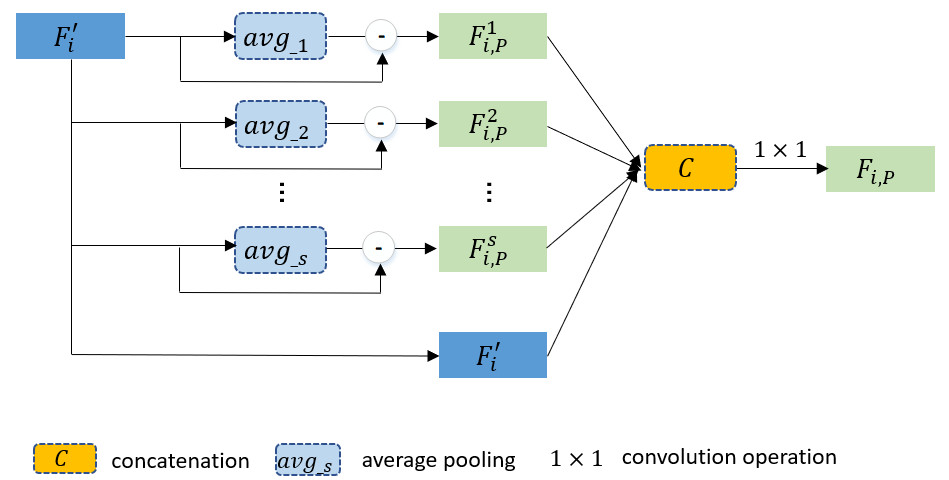}}
	\caption{The structure of the proposed pyramid edge extraction (PEE) module.}
	\label{fig:PEE}
\end{figure}

Edge of lesion region offers important information on the location of target objects. However, the boundary of the lesion area is usually complex and diverse. In order to obtain a robust boundary information supplement, we design a simple and effective pyramid feature extraction scheme for mining multi-granularity edge features in each stage of encoder network. As illustrated in Fig. \ref{fig:PEE}, at first, we use the $1\times 1$ convolution to squeeze the feature maps $F_{i}, i\in\{1,2,3,4\}$ from last residual blocks in each stage of backbone and then employ them as the input for PEE module. It can be defined as follows:

\begin{eqnarray}
&& F'_{i} = \mathscr{F}(F_{i}; \theta_i), \quad \quad i \in \{1,2,3,4\}
\end{eqnarray}
where $F'_{i}$ denotes the reduced feature maps of each residual blocks, $\mathscr{F}$ is the function of $1\times 1$ convolution, and $\theta_i$ indicates the respective parameter. We obtain multiple granularities edge features by subtracting the value of average pooling with different sizes from its local convolutional feature maps. Without losing generality, we define $S$ pooling operations:

\begin{eqnarray}
&& F^{(s)}_{i,\mathcal{P}} = F'_{i} - avg_{\_s}(F'_{i}), \quad s \in \{1,...,S\}
\end{eqnarray}
where $F^{(s)}_{i,\mathcal{P}}$ denotes the edge features of current $i$th stage with the $s$th pooling operation, and $avg_{\_s}$ is the corresponding average pooling operation. In order to integrate the obtained pyramid edge features, we aggregate them with the features of current stage by concatenating them together, and merge them using a $1\times 1$ convolution operation. 

\begin{eqnarray}
&& F_{i,\mathcal{P}} = \mathscr{F}(\mathcal{C}(F^{(1)}_{i,\mathcal{P}},...,F^{(S)}_{i,\mathcal{P}},F'_{i}); \theta_{i,\mathcal{P}})
\end{eqnarray}
where $\mathcal{C}$ refers to the concatenation process. $F_{i,\mathcal{P}}$ is the output feature maps of PEE module at current stage of encoder network. $\theta_{i,\mathcal{P}}$ represents the corresponding parameter. Such a multiple granularities edge features extraction design offers a powerful way to efficiently enhance the representation ability of the corresponding level. By extracting and integrating boundary information of different granularities, the edge features are effectively improved and noise is suppressed. Subsequently, the output maps are fed to the mini multi-task module to promote extraction of finer features. 

\subsection{Mini Multi-Task Learning Module}

\begin{figure}[!t]
	\centerline{\includegraphics[width=0.45\textwidth]{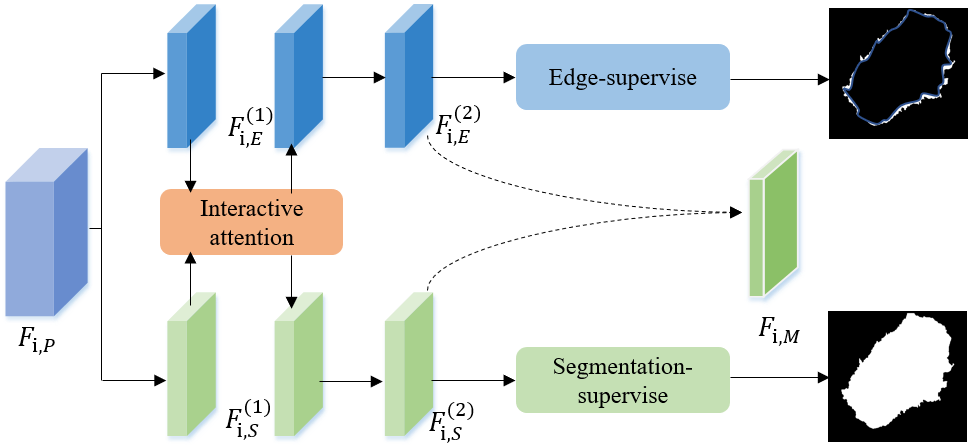}}
	\caption{The design of proposed mini multi-task learning (mini-MTL) module, which consists of two task-specific branches and an interactive attention layer.}
	\label{fig:MTL}
\end{figure}

Naturally, additional knowledge from object edge can help to precisely identify the shape of the target, and the semantic segmentation and edge detection have strong complementary.
Based on this idea, we propose a mini multi-task learning module (mini-MTL) embedded in each stage for jointly learning to segment object masks and detect lesion boundaries without introducing much parameters. The mini-MTL module aims at yielding performance gains by leveraging potential correlations among these related tasks. Fig. \ref{fig:MTL} displays the architecture of our proposed mini-MTL network. Each multi-task networks contains two components: the task-specific branch and an interactive attention layer. Specifically, each branch has two convolutional layers and an upsampling layer. The convolutional layer is used for encoding task-related features and upsampling layer is employed for obtaining the corresponding prediction mask. In $i$th stage, the feature maps $F_{i,\mathcal{P}}$ of PEE module are as the input of two subnetworks for further extracting task-related features simultaneously. 

\begin{eqnarray}
&& F^{(l)}_{i,\mathcal{E}}  = \mathscr{F}(F^{(l-1)}_{i,\mathcal{E}}; \theta^{(l)}_{i,\mathcal{E}}) \nonumber \\
&& F^{(l)}_{i,\mathcal{S}}  = \mathscr{F}(F^{(l-1)}_{i,\mathcal{S}}; \theta^{(l)}_{i,\mathcal{S}}) 
\end{eqnarray}
where $F^{(l)}_{i,\mathcal{E}}$ and $F^{(l)}_{i,\mathcal{S}}$ indicate the feature maps extracted from $l$th convolution layer of edge sub-network and the segmentation sub-network, $l \in \{1,2\}$. In particular, $F^{(0)}_{i,\mathcal{E}}$ and $F^{(0)}_{i,\mathcal{S}}$ denote the feature $F_{i,\mathcal{P}}$ of PEE module. $\mathscr{F}$ is the function of $3\times 3$ convolution with respective parameters $\theta^{(l)}_{i,\mathcal{E}}$ and $\theta^{(l)}_{i,\mathcal{S}}$ respectively. The interactive attention (IA) is designed in the first convolutional layer for mining the interactive information from different tasks. As illustrated in Fig. \ref{fig:IA}, in order to integrate the effective information from other task, we design one simple yet effective interactive attention integration method assigning different weights to various regions. Specifically, take the edge features integration as an example, we first use the sigmoid function to generate a weight mask which indicates the important positions of current edge feature maps $F^{(1)}_{i,\mathcal{E}}$. Then the reverse attention weight is generated by subtracting the weight mask from one. At last, we can selectively send the useful information from the segmentation feature to current edge feature $F^{(1)}_{i,\mathcal{S}}$ by an element-wise product operation. Similarly, the feature integration on segmentation branch is similar to this. Thus, the whole interactive process can be formulated as follows:

\begin{figure}[!t]
	\centerline{\includegraphics[width=0.45\textwidth]{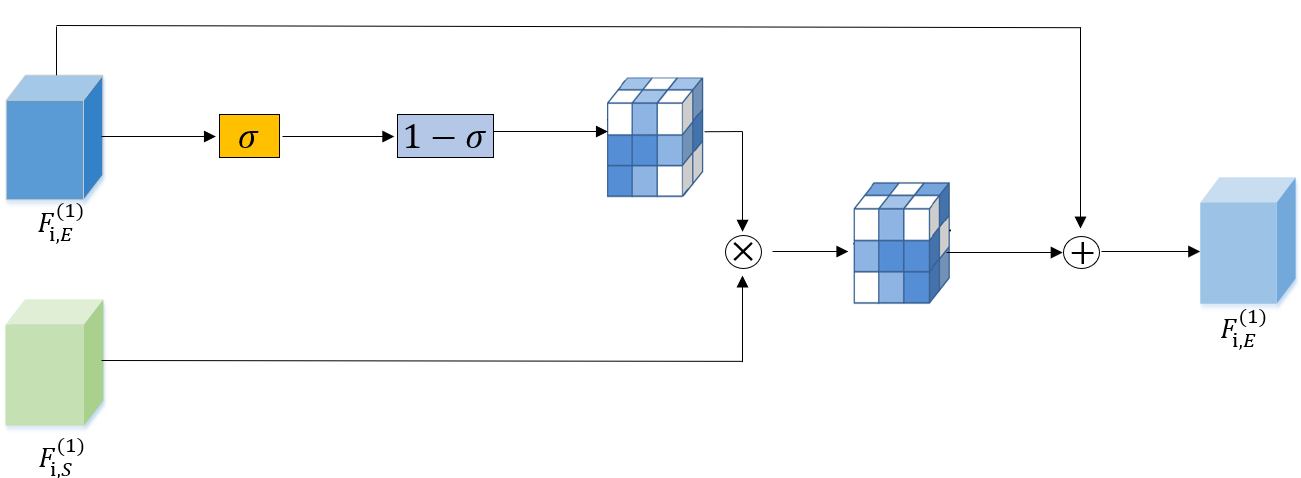}}
	\caption{The illustration of proposed interactive attention method. The figure shows the example of edge features integration by IA.} 
	\label{fig:IA}
\end{figure}

\begin{eqnarray}
&& F^{(1)}_{i,\mathcal{E}} = F^{(1)}_{i,\mathcal{E}} + (1-\sigma(F^{(1)}_{i,\mathcal{E}}))\otimes F^{(1)}_{i,\mathcal{S}} \nonumber\\
&& F^{(1)}_{i,\mathcal{S}} = F^{(1)}_{i,\mathcal{S}} + (1-\sigma(F^{(1)}_{i,\mathcal{S}}))\otimes F^{(1)}_{i,\mathcal{E}}
\end{eqnarray}
where $\sigma$ denotes the sigmoid activation function to produce a filter mask. $\otimes$ represents the element-wise product. The proposed IA module is based on gated mechanism without addition parameters, and in this manner the information from different tasks is delivered effectively by IA module. Moreover, the useful information can be regulated to the right place through attention and useless message can also be suppressed on both the sender and receiver sides simultaneously. After two convolution layers and interaction layer, we get more abundant context representation $F_{i,\mathcal{M}}$ of current stage through aggregating the features from these two tasks:

\begin{eqnarray}
&& F_{i,\mathcal{M}} = \mathscr{F}(\mathcal{C}(F^{(2)}_{i,\mathcal{E}}, F^{(2)}_{i,\mathcal{S}});\theta_{i,\mathcal{M}})
\end{eqnarray}
where $\mathscr{F}$ and $\mathcal{C}$ refer to the $1\times 1$ convolution and concatenation operation respectively. $\theta_{i,\mathcal{M}}$ denotes the parameter. For effectively multi-task learning, we jointly supervise and learn two branches together with IA module in an end-to-end manner. Here, both the edge map and segmentation map are all the binary representation of the outlines of objects and object classes. we expect the output of the edge and segmentation prediction of mini-MTL module to approximate the ground-truth masks respectively (represented as $G_\mathcal{E}$ and $G_\mathcal{S}$) by minimizing the loss:

\begin{eqnarray}
L_{i,\mathcal{E}} = BCE(P_{i,\mathcal{E}}, G_{\mathcal{E}}) \nonumber\\ 
L_{i,\mathcal{S}} = BCE(P_{i,\mathcal{S}}, G_{\mathcal{S}})
\end{eqnarray}
where $BCE(\cdotp,\cdotp)$ means the binary cross-entropy loss function with the following formulation:

\begin{eqnarray}
BCE(P,G) = -\sum_j^{N}(G_ilogP_j + (1 - G_j)log(1-P_j))
\end{eqnarray} 
where $P_j$ and $G_j$ indicate the $j$th pixel of predicted boundary (segmentation) maps $P_{i,\mathcal{E}}$ ($P_{i,\mathcal{S}}$) and ground-truth masks of object boundary (segmentation) $G_{\mathcal{E}}$ ($G_{\mathcal{S}}$), respectively. $N$ represents the number of pixels.
%
Thus the total loss in the $i$th mini-MTL module could be denoted as:

\begin{eqnarray}
L_{i,\mathcal{M}} = L_{i,\mathcal{E}} + L_{i,\mathcal{S}}
\end{eqnarray}
This joint learning helps to preserve the fine details near object boundaries. With this module, BA-Net is able to generate more accurate and better boundary-adherent features at current stage.

\subsection{Cross Features Fusion Module}

\begin{figure}[!t]
	\centerline{\includegraphics[width=0.45\textwidth]{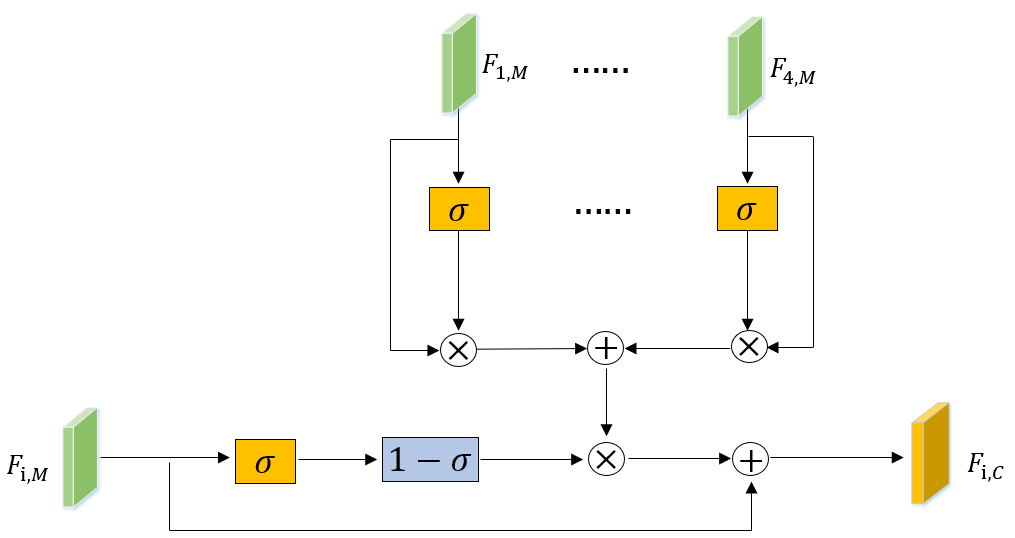}}
	\caption{The design of proposed cross feature fusion (CFF) module.}
	\label{fig:CFF}
\end{figure}

In encoder network, the low-level features are rich in spatial details and the high-level features have more abundant semantic information \cite{r42}. In order to utilize the complementary both spatial structural details and semantic information, we propose cross feature fusion module (CFF) to selectively aggregate features with different levels and refine both high-level and low-level feature maps. As shown in Fig. \ref{fig:CFF},
for $i$th feature maps $F_{i,\mathcal{M}}$, CFF adaptively selects complementary components from multiple input features $F_{j,\mathcal{M}}, j\neq i$ by following attention mechanism:

\begin{eqnarray}
F_{i,\mathcal{C}} = F_{i,\mathcal{M}} + (1-\sigma(F_{i,\mathcal{M}}))\otimes [\sum_{j\neq i} \sigma(F_{j,\mathcal{M}}) \otimes F_{j,\mathcal{M}}]
\end{eqnarray}
where $\sigma$ denotes the sigmoid activation function. $\otimes$ denotes the element-wise product. Therefore, information from different levels is integrated effectively by CFF module and can effectively avoid introducing too much redundant information. Therefore, we can get rich contextual features $F_{i,\mathcal{C}}$ preserving rich details as well as semantic information at each stage through three modules in series which is used for decoder process.  

\subsection{Decoding and Optimization}

By cascaded three modules at each stage of encoder network, richer context and fine-grain features of each stage are encoded. In decoder network, we obtain the decoding features $D_i, i\in\{1,2,3,4\}$ by aggregating the output feature maps $F_{\mathcal{A}}$ from the ASPP module and encoding features of each stage in turn for final segmentation prediction:

\begin{equation}
D_{i}=\left\{
\begin{aligned} 
&\mathscr{F}(\mathcal{C}(D_{i+1}, F_{i,\mathcal{C}}); \theta_{i,D}),  \quad i \in \{1,2,3\}  \\
&\mathscr{F}(\mathcal{C}(F_{\mathcal{A}}, F_{i,\mathcal{C}}); \theta_{i,D}), \quad \quad  i=4
\end{aligned}
\right.
\end{equation}
where $D_{i}$ denotes the decoding features at each stage, $\mathscr{F}$ is the function of $1\times 1$ convolution, and $\theta_{i,D}$ indicates the respective parameter of decoder part. 

The supervision of the whole network adopts the standard binary cross-entropy loss for minimize the error between output from decoder network and ground truth. During the end-to-end training process, the total loss function is defined with joint losses of multi-task module as follows:  

\begin{eqnarray}
min \quad L_D + \sum_i \lambda_i L_{i,\mathcal{M}}
\end{eqnarray}
where $L_D$ denotes the decoder loss and $\lambda_i$ indicates the balance parameters. The $\lambda_i$ is empirically set as $1.0$. The boundary information participates in updating and guiding the generation of final segmentation prediction through loss, which makes the whole network aware of object boundary and refines the result.
 
\section{EXPERIMENTS}

\subsection{Materials}

To evaluate the effectiveness of our method, we conduct experiments on five medical image datasts with various modalities, including dermoscopy images, endoscopic procedures images, X-ray images and retinal fundus images. Details of these datasets are briefly introduced as follows: \textbf{ISIC-2017} \cite{r43} includes a training set with $2000$ annotated dermoscopy images and a total number of $600$ images for testing. The image size ranges from $540 \times 722$ to $4499 \times 6748$. \textbf{Kvasir-SEG} \cite{r44} consists of $1000$ polyp images and their corresponding ground truth polyp masks annotated by expert endoscopists. 
\textbf{CVC-ColonDB} \cite{r52} contains $380$ colonoscopy images coming from $15$ short colonoscopy video sequences with size $574 \times 500$. \textbf{SZ-CXR} \cite{r45,r46} is collected by Shenzhen No.3 Hospital in Shenzhen, China. The dataset contains $566$ X-rays images with respective annotations. \textbf{RIM-ONE-R1} \cite{r53} is composed of $169$ full retinal fundus images which has been annotated by five different experts.

\subsection{Reference Model}

In our work, we compare our proposed BA-Net with six previous state-of-the-art image segmentation methods, including FCN \cite{r25}, U-net \cite{r24}, MultiResUNet \cite{r47}, AG-net \cite{r48}, CE-Net \cite{r23} and Deeplabv3 \cite{r34}. For fair comparisons, the segmentation maps of these methods are generated by the original code released by the authors or directly provided by them. Moreover, all experiments use the same data pre-processing and the predicted segmentation masks are evaluated with the same evaluation metrics.

\subsection{Evaluation Metric}
To quantitively evaluate the performance of the proposed BA-Net, in this paper, five widely accepted metrics for medical image segmentation task are computed as evaluation criteria. They include Dice Similarity Coefficient (DI), Jaccard Index (JA), Accuracy (AC), Sensitivity (SE) and Specificity (SP). The details are defined as follows:

\begin{align}
& DI = \frac{2\cdot TP}{2\cdot TP + FN + FP}, \quad SE = \frac{TP}{TP + FN} \nonumber \\
& JA = \frac{TP}{TP + FN + FP}, \quad \quad SP = \frac{TN} {TN + FP} \nonumber \\
& AC = \frac{TP + TN}{TP + FP + TN + FN}
\end{align}
where $TP$, $TN$, $FP$ and $FN$ represent the number of true positives, true negatives, false positives and false negatives, respectively. They are all defined on the pixel level. Among these metrics, JA mainly reflects the overlapping pixels between estimated and ground-truth masks, which is the most important evaluation metric for segmentation task. In our work, we mainly rank the performance by JA. 

\subsection{Implementation Details}

\textbf{Training setting:} The proposed framework is implemented based on the Pytorch $1.0$ framework and is trained with one NVIDIA TITAN X GPUs. In the model, we use the standard stochastic gradient descent optimizer to train the whole end-to-end network with $0.9$ momentum. The backbone of encoder network is based on ResNet-101 pre-trained on ImageNet. ReLU is chose as the default activation function. We set the initial learning rate as $10^{-4}$ and employ the ``poly" learning rate policy for all experiments similar to \cite{r34}. The learning rate is multiplied by $(1-\frac{iter}{total_iter})^{0.9}$ after each iteration, eventually terminated at $200$ epochs. All the training data is divided into mini-batches for network training, the mini-batch size is set as $8$ during the training stage. In each stage of encoder network, all feature maps are firstly reduced to $128$ channels with convolution operation before PEE module. In our PEE module, we use the $5 \times 5$ and $7 \times 7$ average pooling operation for the first two blocks, and $3 \times 3$ and $5 \times 5$ pooling kernel are employed in last two blocks. All these edge feature maps remain the same size through padding operation. Finally, the output of the decoder is bilinearly interpolated to the same size as the input image directly.

\textbf{Data preprocessing:} In order to enhance diversity of training samples, several common data augmentation and sampling strategies are utilized before training the model. For the sake of fairness, the input preprocessing of all models remains constant. The input images are randomly flipped, rotation and center cropping to perform data augmentation. The scale used in cropping is from $50\%$ to $100\%$ and random rotation in the range of $-10$ to $10$ degrees for original images. Both the horizontal and vertical flip operation with a preset probability are employed in our work. And due to the various size of the image in the datasets, all input images are uniformly resized into a resolution of $256 \times 256$ for training and testing.   

\begin{table*}[!t]
	\renewcommand{\arraystretch}{1.3}
	\caption{Segmentation Performance on four benchmark datasets}
	\label{t1}
	\centering
	\begin{threeparttable}[b]
		\begin{tabular}{L{2.4cm}|C{0.35cm}C{0.3cm}C{0.3cm}C{0.3cm}C{0.35cm}|C{0.35cm}C{0.3cm}C{0.3cm}C{0.3cm}C{0.35cm}|C{0.35cm}C{0.3cm}C{0.3cm}C{0.3cm}C{0.35cm}|C{0.35cm}C{0.3cm}C{0.3cm}C{0.3cm}C{0.35cm}}
			
			\hline
			\hline
			\multirow{2}{*}{\textbf{Methods}} & \multicolumn{5}{c}{\textbf{ISIC-2017}} & \multicolumn{5}{c}{\textbf{Kvasir-SEG}}& \multicolumn{5}{c}{\textbf{CVC-ColonDB}} & \multicolumn{5}{c}{\textbf{SZ-CXR}}\\ 
			\cline{2-21}
			& \textbf{\textit{AC}}  & \textbf{\textit{DI}} & \textbf{\textit{JA}} & \textbf{\textit{SE}} & \textbf{\textit{SP}} 	
	    	& \textbf{\textit{AC}}  & \textbf{\textit{DI}} & \textbf{\textit{JA}} & \textbf{\textit{SE}} & \textbf{\textit{SP}} 	
			& \textbf{\textit{AC}}  & \textbf{\textit{DI}} & \textbf{\textit{JA}} & \textbf{\textit{SE}} & \textbf{\textit{SP}} 
		    & \textbf{\textit{AC}}  & \textbf{\textit{DI}} & \textbf{\textit{JA}} & \textbf{\textit{SE}} & \textbf{\textit{SP}}  \\
			\hline
			FCN \cite{r25}   & 93.9    & 84.1   & 75.2   & 82.2   & 97.0    & 96.7    & 85.9   & 78.9   & 87.5   & 98.1   & 98.5  & 91.4  & 74.0  & 80.0  & 99.6    & 96.7    & 92.9   & 86.9   & 92.0   & 98.2    \\
			U-net \cite{r24} & 93.3    & 85.2   & 76.5   & 84.5   & 97.3   & 96.5    & 85.4   & 78.6   & 86.9   & 98.2   & 98.6   & 83.0   & 75.7   & 81.6   & 99.6 & 98.1    & 96.1   & 92.0   & 95.4   & 99.0     \\
			MultiResUNet \cite{r47}  & 93.6    & 85.2   & 76.8   & 83.9   & 96.8   & 96.8    & 87.0   & 80.5   & 88.5   & 98.2   & 98.5   & 82.8   & 75.6   & 86.9   & 99.1  & 98.1    & 96.0   & 92.4   & 94.6   & \textbf{99.3}  \\
			AG-net \cite{r48}   & 93.5    & 85.3   & 76.9   & 83.5   & \textbf{97.4}  & 97.2    & 88.1   & 81.8   & 88.8   & 98.1    & 98.9   & 84.9   & 76.1   & 83.7   & 99.5  & 98.1    & 96.1   & 92.5   & 95.6   & 99.0\\
			Deeplabv3 \cite{r34}  & 94.0    & 86.4   & 78.3   & 85.9   & 96.9   & 97.3    & 89.2   & 83.5   & 90.4   & 98.2   & 99.2  & 92.0  & 86.4  & 92.0  & 99.5  & 98.0    & 95.8   & 92.2   & 95.3   & 98.9\\
			CE-Net \cite{r23}     & 94.0    & 86.5   & 78.5   & 86.9   & 95.6   & 97.5    & 89.3   & 83.5   & 90.9   & 98.2   & 99.1  & 92.0  & 86.3  & 91.2  & 99.6  & 98.1    & 96.1   & 92.6   & 95.3   & 99.1    \\
			
			\hline 
			\textbf{ours}   & \textbf{94.7}   & \textbf{88.2}   & \textbf{81.0}   & \textbf{89.9}   & 96.4        & \textbf{97.7}    & \textbf{91.1}   & \textbf{86.1}   & \textbf{90.8}   & \textbf{98.6}   & \textbf{99.3}   & \textbf{93.7}   & \textbf{88.4}   & \textbf{93.6}   & \textbf{99.6}   & \textbf{98.2}   & \textbf{96.2}   & \textbf{92.8}  & \textbf{95.8}  & 99.0\\
			\hline
			\hline
		\end{tabular}
	\end{threeparttable}
\end{table*}

\subsection{Comparisons with the State-of-the-Art}
\subsubsection{Dermoscopy image dataset}

Skin lesion segmentation within dermoscopy images is very useful for automated melanoma diagnosis, especially for early protection and treatment. As shown in Table \ref{t1}, our BA-Net attains the highest DI of $88.2\%$ and the best JA of $81.0\%$ than other state-of-the-art methods. As for JA, our approach is noticeably improved from $78.5\%$ to $81.0\%$ on the test set compared to the best competitor CE-Net. In comparison with the results of classical FCN and U-net, our work exceeds them $5.8\%$ and $4.5\%$ on metric JA. Moreover, compared with the latest MultiResUNet and AG-net network, about $4.2\%$ and $4.1\%$ gains are obtained by our method, respectively. The above results suggest that the scheme design of cascaded three modules effectively extracts the richer context and fine-grain features for lesion recognition.


\begin{table}[!t]
	\renewcommand{\arraystretch}{1.3}
	\caption{Segmentation Performance on ISIC-2017 dataset}
	\label{t2}
	\centering
	\begin{threeparttable}[b]
		\begin{tabular}{L{3.3cm}C{0.6cm}C{0.6cm}C{0.6cm}C{0.6cm}C{0.6cm}C{0.6cm}}
			
			\hline
			\hline
			\multirow{2}{*}{\textbf{Methods}} & \multicolumn{5}{c}{\textbf{Averaged evaluation metrics (\%)}} \\ 
			\cline{2-6}
			& \textbf{\textit{AC}}  & \textbf{\textit{DI}} & \textbf{\textit{JA}} & \textbf{\textit{SE}} & \textbf{\textit{SP}} \\
			\hline
			Team-Mt.Sinai (*1)      & 93.4    & 84.9   & 76.5   & 82.5   & 97.5  \\
			Team-NLP LOGIX (*2)     & 93.2    & 84.7   & 76.2   & 82.0   & 97.8  \\
			Team-BMIT (*3)          & 93.4    & 84.4   & 76.0   & 80.2   & \textbf{98.5}  \\
			Team-BMIT (*4)          & 93.4    & 84.2   & 75.8   & 80.1   & 98.4  \\
			Team-RECOD Titans (*5)  & 93.1    & 83.9   & 75.4   & 81.7   & 97.0  \\
			DDN \cite{r36}              & 93.9    & 86.6   & 76.5   & 82.5   & 98.4  \\
			SLSDeep \cite{r50}          & 93.6    & 87.8   & 78.2   & 81.6   & 98.3  \\
			SegAN \cite{r37}            & 94.1    & 86.7   & 78.5   & -      & -     \\
			Chen et al. \cite{r51}      & 94.4    & 86.8   & 78.7   & -      & -     \\
			MB-DCNN \cite{r49}          & -       & 87.8   & 80.4   & -      & -     \\
			\hline 
			\textbf{ours}   & \textbf{94.7}    & \textbf{88.2}   & \textbf{81.0}  & \textbf{89.9}  & 96.4 \\
			\hline
			\hline
		\end{tabular}
		\textbf{note:} The *-number indicates the rank of that method in original ISIC-2017 challenge.
	\end{threeparttable}
\end{table}

We also compare our BA-Net with top five methods submitted to the ISIC-2017 skin lesion segmentation challenge. As shown in Table \ref{t2}, our method achieves the best performance on the benchmark, outperforming the best result of the competition with $4.5\%$ improvement on JA. In addition, we also report the other five recently published methods for skin lesion segmentation on the same test set. It is observed that our BA-Net has obvious advantages compared with the competitive published benchmarks. In particular, BA-Net outperforms the state-of-the-art MB-DCNN by $0.6\%$ on JA. Compared with MB-DCNN using mutual bootstrapping model with three stages, we have achieved outstanding performance in a simple and unified segmentation framework.

\subsubsection{Endoscopic image dataset}

Polyps are predecessors to colorectal cancer and therefore early treatment can help clinicians conduct risk screening and further diagnosis. Our method also achieves the best segmentation performance on both Kvasir-SEG and CVC-ColonDB datasets compared with the above state-of-the-art methods. Table \ref{t1} shows the comparison among them. For the Kvasir-SEG dataset, we divide the Kvasir-SEG including $800$ images for training and $200$ images for testing. Our BA-Net attains JA of $86.1\%$, which outnumbers the second competitor CE-Net and the Deeplabv3 by $2.6\%$. The images in CVC-ColonDB dataset from $15$ different videos, where $304$ images are used for training and $76$ images for testing in our work. From the table, we can find that our method consistently outnumbers other state-of-the-art architectures and shows more performance gain on polyps segmentation. It implies that our method achieves the effective extraction of the same level features and integration of cross level features for improving the segmentation performance.
\subsubsection{X-ray image dataset}

We apply our BA-Net to segment lung organ in 2D X-ray image. In this dataset, we randomly separate the original $566$ images into $452$ training samples and $114$ testing samples. The detailed experiment results are presented in Table \ref{t1}. The BA-Net achieves state-of-the-art performance in most metrics. The proposed method reaches an overall JA of $92.8\%$. In comparison with the results of classical FCN, the JA increases by $5.9\%$. The above results indicate our model with three modules substantially boosts the segmentation performance. 

\subsubsection{Retinal fundus image Dataset} We report the performance of our method for optic disc segmentation from retinal fundus images. The $169$ images from RIM-ONE-R1 dataset are randomly separated to training and test sets with the ratio of $8:2$. The RIM-ONE-R1 dataset contains five independent annotations from five experts. We compare the test results with the corresponding five expert marks, as shown in Table \ref{t3}. From the table, we can observe that our BA-Net achieves the best results both on single evaluation and overall average in JA. It further supports that our proposed PEE, mini-MTL and CFF modules are beneficial for medical image segmentation.

\subsubsection{Qualitative Evaluation} 
\begin{figure*}[!t]
	\centering
	\subfigure{\includegraphics[width=12cm,height=9cm]{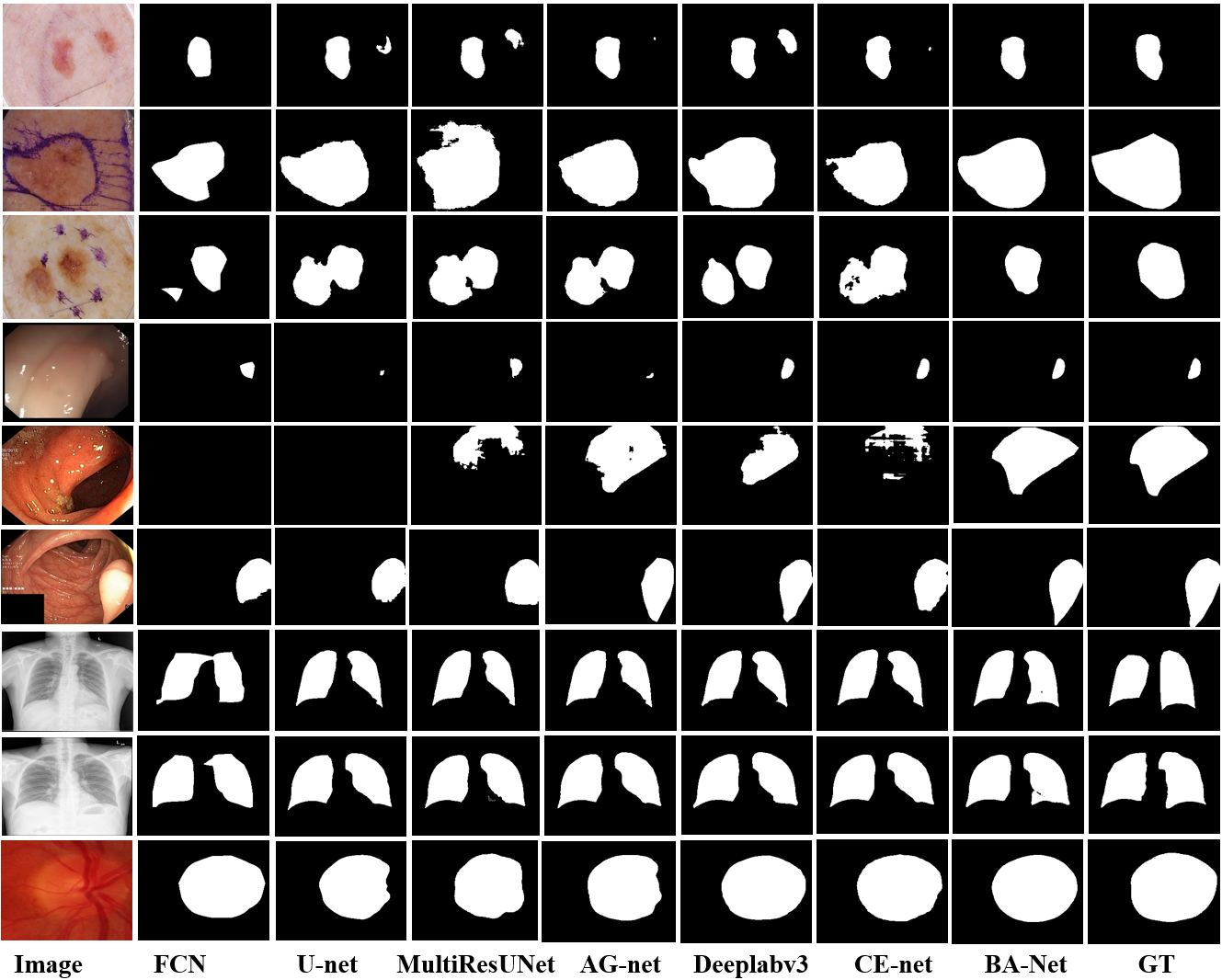}}	
	\caption{Visual comparisons to six state-of-the-art methods on different medical image datasets.}
\label{fig:visual}
\end{figure*}

Fig. \ref{fig:visual} shows some examples of segmentation masks generated by our BA-Net as well as other state-of-the-art methods. We can see that objects can be highlighted well with accurate location and details by the proposed method. Meanwhile, similar background regions and noise are suppressed more thoroughly. From the row of $1$ to $4$ in Fig. \ref{fig:visual}, some methods only segment a part of the lesions and many other non-foreground information also get response. In comparison, our method performs very well. In the $4$th and $5$th rows, when dealing with lesion boundary of similar colors, other methods lack response to target lesions in various degrees, while our method can obtain the clear result. From the row of $6$ to $9$ in Fig. \ref{fig:visual}, we can observe that many methods usually extract coarse region mapping on the whole, but the details of the target area are missing due to noisy backgrounds or lower image contrast. In our method, target areas are more accurately located with the help of effective context representations. Overall, results show that the proposed BA-Net effectively mines the context of same level and uses the underlying correlations among features of distinct levels for capturing more abundant feature embeddings, which indicates the ability of processing the fine structures and rectifying errors.

\begin{table}
	\renewcommand{\arraystretch}{1.3}
	\caption{Segmentation Performance on RIM-ONE-R1 dataset}
	\label{t3}
	\centering
	\begin{threeparttable}[b]
		\begin{tabular}{L{2.5cm}C{0.5cm}C{0.5cm}C{0.5cm}C{0.5cm}C{0.5cm}C{0.5cm}C{1.6cm}}
			
			\hline
			\hline
			\multirow{2}{*}{\textbf{Methods}} & \multicolumn{5}{c}{\textbf{Averaged evaluation metrics--JA (\%)}} & \multirow{2}{*}{\textbf{Overall}}\\ 
			\cline{2-6}
			& \textbf{\textit{E1}}  & \textbf{\textit{E2}} & \textbf{\textit{E3}} & \textbf{\textit{E4}} & \textbf{\textit{E5}}   \\
			\hline
			FCN \cite{r25}                      & 92.1    & 90.6   & 91.1   & 90.1   & 90.9  & 90.9 \\
			U-net \cite{r24}                     & 92.6    & 90.9   & 91.3   & 90.6   & 91.7  & 91.4 \\
			MultiResUNet \cite{r47}             & 93.1    & 91.6   & 92.5   & 91.5   & 92.9  & 92.3 \\
			AG-net \cite{r48}                    & 92.9    & 91.4   & 92.1   & 90.3   & 92.1  & 91.8 \\
			Deeplabv3 \cite{r34}                  & 93.2    & 91.7   & 92.4   & 91.9   & 93.2  & 92.5 \\
			CE-Net \cite{r23}                   & 93.8    & 92.0   & 92.6   & 91.7   & 93.4  & 92.7 \\
			\hline 
			\textbf{ours}   & \textbf{93.9} & \textbf{92.4} & \textbf{92.6}  & \textbf{92.1}  & \textbf{93.7} & \textbf{93.0} \\
			\hline
			\hline
		\end{tabular}
	\textbf{note:} The E-number indicates the different experts.
	\end{threeparttable}
\end{table}

\begin{figure}[!t]
	\centering
	\subfigure{\includegraphics[width=0.48\textwidth]{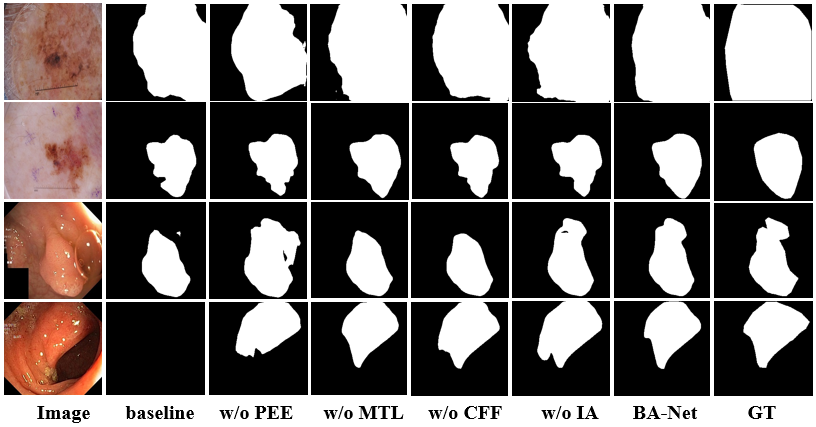}}
	\caption{Segmentation masks inferred with different ablations of our model on ISIC-2017 and Kvasir-SEG dataset. (w/o PEE: without PEE module, w/o MTL: without mini-MTL module, w/o CFF: without CFF module, w/o IA: without IA in mini-MTL module.)}
	\label{fig:visual2}
\end{figure}

\subsection{Ablation Study}
To validate the effectiveness of different components of the BA-Net, several ablation experiments are conducted on the ISIC-2017 and Kvasir-SEG benchmark datasets. The ablations results are shown in Table \ref{t4}.

\subsubsection{Effectiveness of the BA-Net} The proposed BA-Net designs three cascaded module in each stage of encoder network for refining and purifying the context, which achieves the outstanding segmentation performance. To justify the effectiveness, we compare the corresponding results by removing each module (denoted as ``ours w/o PEE'', ``ours w/o MTL'' and ``ours w/o CFF'') respectively.

As shown in Table \ref{t4}, we can observe that the designed three modules greatly improve the segmentation performance compared with ``baseline''. For these two datasets, the JA increases by $3.1\%$ and $2.6\%$ respectively, which indicates the usefulness of these modules for segmentation. After removing the PEE module from our model, the JA declines by $1.0\%$ and $1.6\%$ for both datasets due to the lack of the capability to utilize edge features with multiple granularities. Similarly, without using the mini-MTL module, the ability of the model to make use of the complementarity of boundary information for segmentation is greatly weakened. This is also clearly reflected in the change of two data sets in JA, which decreases by $0.8\%$ and $1.3\%$. Compared with the model only employing the PEE and mini-MTL modules, we can find that the application of CFF yields a $1.1\%$ and $0.7\%$ improvement of JA for these two tasks, respectively. It suggests that the CFF effectively bridges the features of different levels, and effectively integrates them for feature extraction at the current level. In fact, we find that any combination of two modules also brings large gains compared with ``baseline''. When the network utilizes the PEE and mini-MTL module, the context of current level is fully utilized and mined. And the CFF module is employed for further integrating refined contextual information after PEE and mini-MTL modules from cross levels. From Table \ref{t4}, by fully extracting the features of each level and reasonably integrating the information of different stages, our model has achieved excellent results. 

\subsubsection{Effectiveness of the Interactive Attention} In addition, we also investigate the effectiveness of the interactive attention in mini-MTL module. We conduct the experiment removing the IA of mini-MTL module and compare it with our original model. As shown in Table \ref{t4}, for the no-interaction model, it gets the lower score of JA on two datasets and the JA declines by $0.7\%$ and $1.1\%$. It implies that the IA between the edge sub-network and segmentation sub-network plays a great role in generating better representation for final segmentation prediction. As we expect, the IA fully considers the effect of different branches in mini-MTL module and effectively delivers the message to each other.

\subsubsection{Visualization comparison of ablations} Fig. \ref{fig:visual2} shows the segmentation performances with different ablations. We find that ours can better highlight the objects without introducing background regions, and the visualization results of ours are better
than any of w/o PEE, w/o MTL, w/o CFF and w/o IA. Specifically, the baseline model can extract coarse region contour on the whole. But for some images with low contrast and the image with unclear boundary, its segmentation accuracy is not unsatisfactory. From the comparison between module ablations and ours, with the help of PEE, our model obtains richer boundary information guidance, and the mini-MTL module makes full use of the complementarity of boundary information to segmentation. Meanwhile, the CFF module provides cross-level feature interaction, which enables the model to capture more overall context at each stage. In addition, the IA in mini-MTL enhances the delivery of effective information from two branches. This further demonstrates that the introduction of three modules designed in our work effectively enhances the segmentation performance.

\begin{table}[htbp]
	\centering
	\begin{threeparttable}[t]
		\caption{\label{t4} Ablation analysis for the proposed network on ISIC-2017 and Kvasir-SEG datasets.}
		\begin{tabular}{L{2.5cm}|C{1cm}C{1cm}|C{1cm}C{1cm}}
			\hline
			\hline
			\multirow{2}{*}{\textbf{Methods}} & \multicolumn{2}{c}{\textbf{ISIC-2017}} & \multicolumn{2}{c}{\textbf{Kvasir-SEG}} \\
			\cline{2-5}
			& \textbf{\textit{DI}} & \textbf{\textit{JA}} & \textbf{\textit{DI}} & \textbf{\textit{JA}}\\
			\hline
			baseline            & 86.1  & 77.9   & 89.2  & 83.5 \\
			\hline
            ours w/o PEE      & 87.6  & 80.0   & 90.1  & 84.5 \\
            ours w/o MTL      & 87.7  & 80.2   & 90.4  & 84.8 \\
            ours w/o CFF      & 87.3  & 79.9   & 90.8  & 85.4 \\
			ours w/o IA       & 87.8  & 80.3   & 90.6  & 85.0 \\
			\hline
			\textbf{ours}     & \textbf{88.2}  & \textbf{81.0}   & \textbf{91.1}  & \textbf{86.1} \\
			\hline
			\hline
			
		\end{tabular}
	\end{threeparttable}
\end{table}

\section{Conclusion}

In this paper, we present a boundary-aware context neural network, which produces richer contextual information for medical image segmentation. In our work, three cascaded modules are proposed. In each stage of encoder network, the pyramid edge feature extraction module extracts multiple granularities edge features, and then the mini multi-task learning module effectively complements and enriches the context from the boundary detection branch through multi-task learning and designed interactive attention method. Finally, the network adaptively integrates the feature maps from different levels by the cross feature fusion module. By cascaded three modules, richer context and fine-grain features of each stage are obtained. Extensive comparative evaluations on five publicly available datasets are implemented, which has validated the superiority of the proposed method. Based on the outstanding performance of our work, we will extend our BA-Net to support 3D medical image segmentation tasks in the future.


\ifCLASSOPTIONcaptionsoff
  \newpage
\fi

\end{document}